\documentclass[preprint,pra,aps,showpacs,preprintnumbers,amsmath,amssymb,eqsecnum]{revtex4}

\usepackage{graphicx}
\usepackage{dcolumn}
\usepackage{bm}
\usepackage{verbatim}
\usepackage{epsfig}
\usepackage{epstopdf}

\newcommand\beq{\begin{equation}}
\newcommand\eeq{\end{equation}}
\newcommand\bea{\begin{eqnarray}}
\newcommand\eea{\end{eqnarray}}

\def\Q1{{\hat Q(t_1)}}
\def\Q2{{\hat Q(t_2)}}

\def\id{{1 \!\! 1 }}
\def\half{\frac {1} {2}}

\def\x0{{{\bf x}_0}}

\begin{document}


\title{Leggett-Garg Correlation Functions from a Non-Invasive Velocity Measurement Continuous in Time}

\author{J.J.Halliwell}%
\email{j.halliwell@imperial.ac.uk}

\affiliation{Blackett Laboratory \\ Imperial College \\ London SW7 2BZ \\ UK }



\begin{abstract}
In the Leggett-Garg approach to testing macrorealism, the two-time correlation functions, which are normally obtained by sequential measurements of a dichotomic variable $Q$, need to be measured in a non-invasive way in order to exclude certain types of alternative classical explanation. Here, it is shown, for a class of macrorealistic theories,
that the correlation functions are readily expressed in terms of a time integral of the velocity corresponding to $Q$ and that this expression can be determined from a single final-time measurement of an auxiliary system in continual weak interaction with the primary system. The protocol has the form of a ``waiting detector'' which clicks only when $Q$ changes sign. It shares features with both ideal negative measurements and weak measurements and we argue that it is essentially non-invasive, under certain reasonable assumptions.
We show that the non-invasiveness persists to a quantum model of the process.
\end{abstract}








\maketitle

\section{Introduction}

The Leggett-Garg (LG) inequalities were proposed in order to test whether certain types of realist theories could explain the observable data in the description of macroscopic systems \cite{LG1,L1}. In this approach, sequential measurements are made on a dichotomic variable $Q$ at three or more times, from which temporal correlation functions of the form
\beq
C_{12} = \langle Q(t_1) Q(t_2) \rangle,
\label{corr}
\eeq
are obtained.
Under the assumptions that these variables take definite values (macrorealism per se), that they can be measured without disturbing the future evolution of the system (non-invasive measurability), and that future measurements cannot affect the past, it can be shown that the correlation functions obey the Leggett-Garg inequalities \cite{LG1,L1}, which for three times have the form
\bea
1 + C_{12} + C_{23} + C_{13} & \ge & 0,
\\
1 - C_{12} - C_{23} + C_{13} & \ge & 0,
\label{LG2}
\\
1 + C_{12} - C_{23} - C_{13} & \ge & 0,
\\
1 - C_{12} + C_{23} - C_{13} & \ge & 0.
\label{LG4}
\eea
These inequalities are respected by certain interesting classes of macrorealistic theories, when properly measured.  (For a discussion of the current status of this field see the useful review Ref.\cite{ELN} and the critique Ref.\cite{MaTi}).

The Leggett-Garg inequalities can be violated by quantum mechanics. Consider for example the commonly-studied spin system in which $H = \half \omega \sigma_x $ and $Q = \sigma_z $. The equations of motion have solution
\beq
\sigma_z (t_2) = \cos  \omega (t_2 - t_1) \  \sigma_z (t_1) + \sin \omega (t_2 -t_1) \ \sigma_y (t_1).
\eeq
The correlation function is given by
\bea
C_{12} &=&  \half \langle \psi | \hat Q(t_1) \hat Q(t_2) + \hat Q(t_2) \hat Q (t_1) | \psi \rangle,
\label{corrq}
\\
&=& \half \langle \psi |  \{ \sigma_z (t_1), \sigma_z (t_2) \} | \psi \rangle,
\nonumber \\
&=& \cos \omega (t_2 -t_1),
\eea
and is independent of the initial state. If we choose the time intervals to be equally spaced, $ t_1 - t_2 = t = t_3 - t_2$, we find that the LG inequality Eq.(\ref{LG2}), for example, reads
\beq
1 - 2 \cos \omega t + \cos 2 \omega t \ge 0.
\label{LG5}
\eeq
This is violated for $ 0 < \omega t < \pi / 2 $, with a maximal violation in which the left-hand side takes value $ - \half $ at $\omega t = \pi / 3$.

Such violations have been confirmed in numerous experimental tests \cite{Pal,Jor,Knee,Rob,Geo}. However, what is important in such tests is the requirement of non-invasive measurability (NIM), analogous to the locality requirement in Bell inequality tests.
This is crucial since if invasive one can argue that it is the disturbance of the measurement that produced the particular form of the correlation function \cite{deco} and indeed there are specific models that show exactly how the quantum correlation function can be classically replicated \cite{Mon,Ye1,Guh,KS}.

The NIM requirement is demanding to implement and not many experiments have done so in a fully satisfactory way
 \cite{Knee,Rob,Geo,KBLL,KneeMRtest}. Leggett and Garg in their original proposal suggested that non-invasiveness is accomplished using an ideal negative measurement, in which the detector measuring $Q$ at the first time couples to only one of its values, $Q=+1$ say, and if the detector does not register, it is deduced that the system was in state $Q=-1$ at the first time. This is reasonable for the macrorealistic theories being put to the test since they are essentially classical in nature, but this procedure would in general be invasive for a quantum mechanical system, since the wave function still collapses \cite{Dicke}. 
Weak measurements \cite{weak,weak2,Wang} have also been used to measure the correlation functions, in an arguably minimally invasive way \cite{Pal,Jor,Rus,Gog}, although it remains a matter of debate as to whether they fully meet the NIM requirement \cite{deco,ELN}. Another approach to NIM is the ``stationarity'' requirement \cite{stat} which has also met with some criticism \cite{ELN}.

A significant feature of the NIM requirement is that many experimental tests implementing it rely on an argument for non-invasiveness that is appropriate to macrorealistic systems but will not in general hold up in quantum mechanics. Since we expect experimental devices to adhere to the laws of quantum mechanics this means that non-invasiveness cannot be checked experimentally in most protocols. 
There are, however, recent attempts to address this \cite{KBLL,KneeMRtest} which brings NIM into the domain of experiemental checks (and see also the earlier discussions \cite{L1,deco}) and it is clearly of interest to develop this further. Indeed, we note that some protocols can be made quantum-mechanically non-invasive for special choices of initial state, for example, in an ideal negative measurement,
taking the state to be an eigenstate of $\hat Q$  at the first time.

Although many experimental approaches to date can reasonably claim success in implementing the LG programme, the difficulty in experimentally implementing some protocols together with the question marks (in some cases) around non-invasiveness indicates that it remains of interest to search for alternative approaches which may be easier to implement and have a clearer or at least different story in terms of NIM. The purpose of this paper is to present a different type of protocol.

We first note that what is perhaps common to most if not all approaches to date is the focus on a pair of measurements acting at two successive times and the origin of potential invasiveness is the fact that the earlier measurement could affect the result of the later one. Hence in the search for different approaches which may give an alternative perspective on NIM, it would be of interest to find a protocol which gets away from this feature. To this end, note that the correlation function we seek may be written
\beq
C_{12} = p(S) - p(D)
\eeq
where $p(S)$ denotes the probability that $Q(t)$ takes the same value at $t_1$ and $t_2$ and $p(D)$ denotes the probabilty that they take different values. (This is spelled out in more detail in Refs.\cite{HalQ,Halprev}). Hence the correlation function depends only on whether $Q(t)$ has the same or opposite signs at the initial and final times and not on the specific initial or final value of $Q(t)$.
This suggests that the correlation function could be determined by examining whether $Q(t)$ changes sign in each run, or not, during the given time interval. Of course it could change sign more than once during the time interval, but this is a question of timescales which can be addressed in specific models, and we will in fact see that there is a regime in which it is reasonable to assume no more than one sign change.

Given that only the sign changes are important, and not the actual value of $Q$ itself, we can then look for a detection scheme which measures whether a change takes place at {\it any} time during $[t_1,t_2]$. This involves a ``waiting detector'', that clicks or not during the given time interval, depending on whether $Q$ changes sign at any time. (An analogous approach is sometimes used in the arrival time problem \cite{Wait}).
In this paper a simple protocol which has precisely this property is presented.
It will be argued that this protocol is essentially
non-invasive and furthermore the non-invasiveness persists to the quantum level, which suggests
the possibility of confirming non-invasiveness through an experimental check.

The protocol bears some resemblance to a protocol presented in an earlier paper by the present author
in which the two-time histories of the system were measured directly in two different ways, firstly, 
by an ancilla coupled to the primary system with two CNOT gates \cite{Halprev}, and secondly, using a single final time measurement that was argued to be correlated with the two-time history.
The protocol in the present paper is simpler but similar in spirit in that it again effectively measures the histories of the system.

To be clear, the protocol given here is a general theoretical sketch of a possible experiment. The construction of specific  experimental tests along the lines presented here is beyond the scope of this paper and no claims of experimental feasibility are made.
However,  we note that the protocol considered here bears some resemblance to the recent experiments described in Refs.\cite{Knee,Rob,KBLL}, in which measurements of simple systems using an ancilla are considered.

\section{The Protocol}

\subsection{Some observations}

We start from the simple observation observation that
\beq
\langle [Q(t_2) - Q (t_1) ]^2 \rangle  = 2 (1 -  C_{12}).
\eeq
This holds in both the classical and quantum theory, with correlation functions given by Eqs.(\ref{corr}) and (\ref{corrq}) respectively. It suggests that the correlation function can be determined from a measurement of the quantity $Q(t_2) - Q(t_1) $.

Classically, $Q(t_2) - Q(t_1)$ can take the values $0$ or $\pm 2$. By contrast, in the quantum theory, its possible values are determined by looking at the spectrum of the corresponding operator. In the case of the simple spin model above this is
\bea
\hat Q(t_2) - \hat Q(t_1)
 &=&
\left( \cos  \omega (t_2 - t_1) - 1 \right) \  \sigma_z (t_1) + \sin \omega (t_2 -t_1) \ \sigma_y (t_1),
\nonumber \\
&\equiv&
{\bf a} (t_1,t_2) \cdot {\overrightarrow \sigma}.
\eea
Quantities of this type can be measured by measuring spin along the direction specified by the vector ${\bf a} (t_1,t_2)$, and will take values
$ \pm |{ \bf a} |$, where $ {\bf a}^2 = 2 ( 1 - C_{12} )$. Such a measurement would be different for different time intervals and involves knowledge of the quantum dynamics to determine the vector ${\bf a}$
so cannot obviously be expressed in terms a macrorealist could work with.
Note, however, that these properties have the following consequence: if the system is in an eigenstate of $\hat Q(t_2) - \hat Q(t_1)$, measuring this quantity yields one of its eigenvalues and does not disturb the state.
This suggests that the correlation function can be measured in a single non-invasive measurement.
Note also, that there is no immediate analogue in the quantum case of the classical case in which $ Q(t_2) =Q (t_1) $. The closest one can get, perhaps, is the case in which $\langle \hat Q(t_2) \rangle = \langle  \hat Q(t_1) \rangle $, which is achieved in any eigenstate of the operator ${\bf b} \cdot
\overrightarrow \sigma $, where ${\bf b}$ is any vector orthogonal to ${\bf a}$.
For example, $\sigma_x$ eigenstates do the job.

\subsection{A macrorealistic formulation of the protocol}

We now seek a way of carrying out a measurement of  $ Q(t_2) - Q (t_1) $ describable in more macrorealistic terms.
The discussion with be given for a general dichotomic variable $Q$, with occasional reference to the simple spin model above. Also, the underlying hidden variable theories we have in mind here are of the GRW type \cite{GRW}, consistent with the critique Ref.\cite{MaTi}.

It is normally assumed that the macrorealistic theory we are testing has a variable $Q$ which takes definite values at every time and has some sort of dynamics, which in general will be stochastic. However, in the present approach we will assume, in addition, that the theory has a velocity variable $ v(t)$ such that $ \dot Q(t) = v (t) $, which can be measured.
If such a velocity variable exists, it then follows that
\beq
Q(t_2) - Q(t_1) = \int_{t_1}^{t_2} dt \ v(t).
\label{vel}
\eeq
The problem of measuring the correlation function is therefore replaced with that of measuring the time-averaged velocity.

Clearly in the quantum case for spin variables, $v$ exists and will simply be a combination of Pauli matrices, just like $Q$.
For example, in the simple spin model above, the velocity operator is $\dot \sigma_z = \omega \sigma_y $.
In a classical stochastic model, $Q$ takes values $\pm 1$, and the definition of velocity is more subtle.
For example, in many hidden variable models for spin systems, $Q$ has the form
\beq
Q(t) = {\rm sign} ( {\bf n} \cdot {\bf x} (t)),
\eeq
where ${\bf x} (t)$ is a unit vector belong to a stochastic ensemble and evolving in time under rotations on the sphere, and ${\bf n}$ is a fixed unit vector \cite{Mon}. Differentiating,
\beq
\dot Q(t) = 2\ {\bf n} \cdot \dot {\bf x} (t) \ \delta  ( {\bf n} \cdot {\bf x} (t))
\label{Qdot}
\eeq
Suppose we now take ${\bf n}$ to lie in the $z$-direction, so that $Q$ is the classical counterpart of $\sigma_z$ in the quantum case.
If the classical evolution is that corresponding to the quantum evolution with $H $ proportional to $\sigma_x$, the vectors ${\bf x}(t)$ rotate on circles of constant $x$,
\beq
{\bf x}(t) = ( x_0, r_0 \cos \omega (t-t_0), r_0 \sin \omega (t-t_0) ),
\eeq
where $x_0^2 + r_0^2 = 1$ and $t_0$ labels the members of the stochastic distribution of vectors. It is then easy to show that, when the $\delta$-function constraint of Eq.(\ref{Qdot}) holds,
\beq
 {\bf n} \cdot \dot {\bf x} (t) = \omega \ {\rm sign} (  {\bf m} \cdot {\bf x} (t)),
\label{Sigy}
\eeq
where ${\bf m}$ is a unit vector lying in the $y$-direction. Assuming a uniform distribution of stochastic vectors ${\bf x}(t)$, we can then average the velocity over the ensemble with the result
\beq
\langle \dot Q(t) \rangle =  \omega \langle  {\rm sign} (  {\bf m} \cdot {\bf x} (t)) \rangle.
\eeq
This means that the classical model replicates the quantum result $\dot \sigma_z = \omega \sigma_y$ in the average.
At the level of individual stochastic trajectories,
it is natural to regularize the $\delta$-function in Eq.(\ref{Qdot}), for example by discretizing the time, and this equation then says that $\dot Q(t)$ is zero except when $ {\bf n} \cdot {\bf x} (t)$ is close to zero, in which case the velocity is
proportional to Eq.(\ref{Sigy}), so there is still a link to the quantum result,
$\dot \sigma_z = \omega \sigma_y$. These arguments indicate that it is reasonable to suppose in a typical hidden variable model that a velocity exists with the requisite properties.

The time-averaged velocity can be measured quite easily using a weak coupling to another system continuous in time.
To give a very simple example, suppose we couple the primary system to a point particle with momentum $p$ and position $q$, using the total Hamiltonian
\beq
H = H_S + \lambda v  p + \frac {p^2} {2m}
\eeq
where $\lambda$ is a small constant. The equation of motion of $q$ is
\beq
\dot q = \frac {p} {m} + \lambda v
\eeq
and $p$ is constant, so the solution is
\beq
q(t_2) = \frac {p (t_2-t_1)} {m} + \lambda \int_{t_1}^{t_2} dt \ v(t),
\eeq
where,
to leading order for small $\lambda$, $v(t)$ is the time-evolved velocity determined by the unperturbed system Hamiltonian $H_S$. Hence the time-averaged velocity is determined from the shift in $q$ at the final time.
One can easily find other similar models which effect a measurement of the time-averaged velocity in this way, for example a coupling to an auxiliary system (ancilla) of the form $ v H_A$, where $H_A$ is the auxiliary system Hamiltonian, so that the auxiliary system dynamics simply switches on when $v$ is non-zero.
Furthermore, it is easy to see how this works in quantum theory. The evolution of the coupled system is determined by the $S$-matrix,
\beq
S (t_1, t_2) = T \exp \left( - i \lambda \int_{t_1}^{t_2} dt \ \hat v (t) \otimes \hat H_A  \right),
\label{Smat}
\eeq
where $T$ us the usual time-ordering operator,
from which we see that the ancilla responds to the time-average of the velocity, to leading order in $\lambda$.

This measurement can be regarded as a weak measurement continuous in time if $\lambda$ is sufficiently small. It may therefore be subject to some of the criticisms of the use of weak measurements in this context \cite{ELN,deco}.
However, we shall now argue that the protocol does much better than standard weak measurements in terms of meeting the NIM requirement.

In general, during a given time interval, $Q(t)$ will jump many times between the values $\pm 1$. There is an interaction with the ancilla each time it changes sign but no interaction at any other times.
The object we are interested in, $Q(t_2) - Q(t_1)$, can take the three possible values $\pm 2$ or zero. 
If it is zero, this in general means that either  $Q(t)$  did not change sign at all, so there was no interaction, or that it jumped value any even number of times.  If it is non-zero, it means that $Q(t)$ changed sign an odd number of times.

However, let us suppose that the total timescale is sufficiently short that  $Q(t)$ will only have time to make either one jump in value, or no jumps. In particular, if it makes no jumps, then the velocity is zero during the time interval, there is no interaction and the ancilla does not change at all during that time. This is clearly promising in terms of finding a non-invasive measurement. However, if we find that the ancilla is unchanged from its initial state at the final time, it means that either, $Q(t)$ did not change sign at all, or, that it did change sign once (and thus interact with the ancilla), but the back action of the ancilla on the primary system caused $Q(t)$ to change sign again, causing a further interaction with the ancilla. This second possibily has small probability for small $\lambda$ and can in fact be estimated in a specific model (and can also be ruled out using an irreversible ancilla, as we see shortly). This means that the situations in which the ancilla is unchanged are, to a good approximation for small $\lambda$, non-invasive determinations of the probability of $Q(t)$ not changing sign. This is therefore a reasonably close analogue of an ideal negative measurement, extended to continual measurement over a time interval, subject to the assumption of sufficiently short timescale and weak interaction. Furthermore, if the ancilla is unchanged,
a final measurement on the primary system can be made to check if it has been disturbed or not during.

Consider now the situation regarding invasiveness in the case where $Q(t)$ changes sign just once. Here, there will be an interaction with the ancilla at the moment when $Q(t)$ change sign (and only then), which could be at any time between the initial and final time. The question then is whether this interaction can possibly affect the future evolution of the primary system. It could, as indicated above, through the back action on the primary system which may cause $Q(t)$ to change sign again and hence influence the final ancilla state. As stated we expect this effect to be small for small $\lambda$. 
However, there is a more elaborate and interesting possibility, which is to choose the ancilla to be effectively irreversible, for example, by using a large system in a metastable state which undergoes an irreversible change when triggered. 
This means that the ancilla does not have the possibility of returning to its ``no detect" state. Of course, the future evolution of $Q(t)$ after detection is changed by the interaction but we are working in a regime in which $Q(t)$ undergoes only one (or no) sign change, so having already been detected making that one sign change,  its future evolution is irrelevant.
Hence this situation is also non-invasive since 
there is no sense in which any future measurements could be affected by the interaction.

In brief, this ``waiting detector" protocol gives an approximately non-invasive account of the measurement of whether $Q$ changes sign or not, under an assumption of sufficiently short timescale together with an assumption of sufficiently weak interaction or an irreversible ancilla.

\subsection{The short timescale assumption}

The requirement of sufficiently short timescale is a plausible one. As noted above, in the hidden variable model above $Q(t) $ is determined by the sign of an ensemble of functions ${\bf n} \cdot {\bf x} (t)$, which behave like
$ \sin \omega (t-t_0) $, for a distribution of initial times $t_0$. If we choose $Q=+1$ initially, this restricts to vectors ${\bf x} (t)$ in the hemisphere defined by ${\bf n}$. Under subsequent time evolution the ${\bf x}(t)$ vectors move into the opposite hemisphere but none of them move back to the original hemisphere as long as the total time interval is less than $ \pi / \omega $. The largest time interval is set by the correlation function $C_{13}$ in the LG inequality Eq.(\ref{LG5}), for example. This interval is $2t$, with the equal time-spacing chosen above, hence there will be only one sign change in this hidden variable model if
$ \omega t \le \pi / 2$. This is a sufficiently large time range to find significant violations of the LG inequalities and in particular include the maximal violation.

One could also contemplate carrying out experimental checks of the typical number of sign changes made by $Q$.
We would fully expect such checks to conform to the laws of quantum mechanics, hence
it is of interest to estimate the fraction of histories making two sign changes of $Q$ using a quantum model. 
If we take $H = \omega \sigma_x / 2 $ and denote the $Q= \sigma_z$ eigenstates by $ | \pm \rangle $, we have
\beq
e^{ - i H t } | + \rangle= \cos \left( \frac { \omega t} {2} \right) | + \rangle - i  \sin \left( \frac { \omega t} {2} \right) | - \rangle,
\eeq
and similarly for the $ |- \rangle $ state. The probability that $Q$ takes values $+1$, $-1$, $+1$ at times $0$, $t$ and $2t$, is then
\bea
p(+,-,+) &=&  \left| \langle + | e^{ - i H t } | - \rangle \right|^2 \  \left| \langle - | e^{ - i H t } | + \rangle \right|^2
\nonumber \\
&=& \sin^4 \left( \frac { \omega t} {2} \right).
\eea
Similarly, for the probability at $Q$ takes values $+1$ at all three times, we have
\beq
p(+,+,+) = \cos^4 \left( \frac { \omega t} {2} \right).
\eeq
It is then useful to define the ratio of paths with two sign changes to paths with none:
\bea
\xi &=& \frac { p(+,-,+) } { p(+,+,+) }
\nonumber \\
&=& \tan ^4  \left( \frac { \omega t} {2} \right).
\label{xi}
\eea
This quantity grows very slowly from zero for small $t$ and at the time $\omega t = \pi  / 3$ where the LG inequality Eq.(\ref{LG5}) has its maximal violation of $- \half $, we have $ \xi = 1/9$. This is small, but perhaps not insignificant. However, at the slightly earlier time $ \omega t = 4 \pi/15 $, $\xi$ drops down to $\xi \approx 0.04$ but the LG violation remains significant at $ -0.44$.

These arguments show that the assumption of only one sign change at sufficiently short time intervals is a reasonable one in both classical and quantum models. However, there will in general
be a small fraction of the ensemble which will have two sign changes.
This will mean that a ``no-detection'' result in the auxiliary system includes some histories in which the velocity was non-zero and so some interaction took place thereby providing opportunity for alternative classical explanations in that fraction of the histories.
As long as the parts of the correlation functions due to two sign changes in $Q$ are sufficiently small (and as long as those parts do not make a significantly ``adversarial'' contribution)
macrorealism can still be tested by demanding that the violations of the LG inequalities are sufficiently large to outstrip classical explanation. Precise modelling of arguments along these lines was given in Refs.\cite{Knee,Rob,KBLL,Halprev}.

\subsection{Comments on weak measurements}

Experimental measurements of the fraction of times $Q$ changes sign or not allow us to determine the correlation function through the formulae,
\bea
p(S) = \half ( 1 + C_{12} ),
\label{same}
\\
p(D) = \half( 1 - C_{12} ).
\label{diff}
\eea
However, due to the weakness of the coupling to the detector, the measurements will not in fact determine $p(S)$ exactly, but determine an expression modified by terms representing the inefficiency of the detector (as we shall see in an explicit model). This is because $\lambda$ represents the rate of transition from the undetected to detected state, but for weak couplings, only a partial transition is made.

Weak measurements of $Q(t)$ continuous in time have been used previously in LG inequality tests \cite{Rus} and indeed the first experimental test was of this type \cite{Pal}. However, as noted in Refs.\cite{ELN,deco}, the weakness of the measurement alone does not ensure non-invasiveness although it does allow the NIM condition to be stated differently. Such a condition was given in Ref.\cite{Rus}, although it has been noted that it will not in general be satisfied in a quantum model \cite{deco}.

The present protocol, although a continuous in time weak measurement, is different in that there is an argument for non-invasiveness, up to small back action effects, 
and one can check experimentally, in the case of no sign changes of $Q$, that the primary system state is unaffected by the measurement. We will see that this non-invasiveness persists at the quantum level with a suitable choice of initial state.

We also note that a weak measurement protocol for the quasi-probability $q(s_1,s_2)$ built from the correlation function (and the two averages $\langle Q(t_1) \rangle$, $\langle Q(t_2) \rangle $) was proposed in Ref.\cite{HalQ}. Here the NIM requirement consists of the no-signaling in time condition \cite{KoBr}, generalized to quasi-probabilities and this condition is in fact satisfied in the quantum case.

\subsection{Relation to quantum backflow}

To close this Section, and as an aside, we note that the analogy with the arrival time problem together with the relationship Eq.(\ref{vel}), suggests a comparison with the quantum-mechanical current for a point particle and its relationship with the probabilities for remaining in the positive or negative $x$-axis. Furthermore, it is known that this current can exhibit ``backflow", in which the current has the opposite sign to the momentum of the state \cite{Back}.
One might speculate that this non-classical phenomenon, or at least its analogue in the systems considered here, is responsible for the quantum-mechanical values of the correlation functions which violate the LG inequalities.
Backflow requires that the underlying Wigner function is negative somewhere. By contrast, it was shown in Ref.\cite{Halprev} that the properties of the systems considered here at two times may be described by a quasi-probability which is positive in some regimes, but the LG inequalities, which refer to three or more times, can still be violated in those regimes. This indicates that backflow and LG inequality violation are two different types of quantum phenomena, the first apparent at just two times, the second not apparent until three or more times.
Nevertheless it remains of interest to explore any possible connections more thoroughly

\section{A Quantum Model}

We now consider the implementation of the protocol in a specific quantum model, consisting of the simple spin system described above, with its velocity $ \omega \sigma_y$ coupled to a two-state auxiliary system initially in the state $|0\rangle$, which switches to the state $|1\rangle$ when interacted with. (We will not however attempt to model an irreversible detector here, although such a model for the analogous arrival time problem has been constructed \cite{HalDet}). 
The total Hamiltonian is
\beq
H= \frac {\omega} {2} \sigma_x \otimes \id + \lambda \omega \sigma_y \otimes \left( | 0 \rangle \langle 1 | + | 1 \rangle \langle 0 | \right).
\label{ham}
\eeq
We note that $H^2 = (\Omega^2 / 4) \id $, where $ \Omega = \omega \sqrt{ 1 + 4 \lambda^2} $, from which it is easily shown that the unitary time evolution operator is
\beq
e^{ - i H t} = \cos \left( \frac {\Omega t }{2}\right) \id -  \frac {2i} { \Omega} \sin \left( \frac{ \Omega t }{2}\right) H.
\eeq
The total system state at time $t$ is,
\bea
|\Psi_t \rangle &=&  e^{-iHt} | \psi  \rangle \otimes |0 \rangle
\nonumber \\
&=&  \hat A_0 (t) | \psi \rangle \otimes |0 \rangle + \hat A_1(t) | \psi \rangle \otimes |1 \rangle,
\label{Psit}
\eea
where
\bea
\hat A_0 (t) &=& \cos \left( \frac {\Omega t }{2}\right) \id -\frac {i \omega } { \Omega} \sin \left( \frac{ \Omega t }{2}\right) \sigma_x,
\\
\hat A_1(t) &=&- \frac {2i \lambda \omega } { \Omega} \sin \left( \frac{ \Omega t }{2}\right) \sigma_y.
\eea
Note that
\beq
\hat A_0^\dag (t) \hat A_0 (t) + \hat A_1^\dag (t) \hat A_1 (t) = 1.
\eeq
and also that each term in this expression is proportional to the identity operator.

The probability of finding the ancilla in the detected state is
\bea
p(1) &=& \langle \psi | \hat A_1^2 (t) | \psi \rangle
\nonumber \\
&=& \frac {2 \lambda^2 \omega^2 } {\Omega^2} (1 - \cos \Omega t ).
\eea
For small $\lambda$, $ \Omega \approx \omega $ and we find
\beq
p(1) \approx 2 \lambda^2 ( 1 - \cos \omega t).
\eeq
This therefore agrees with the result expected (from the $S$-matrix),
\bea
p(1) &=& \lambda^2 \langle ( \hat Q(t_2) - \hat Q(t_1) )^2 \rangle,
\nonumber \\
&=& 2 \lambda^2 ( 1 - C_{12} ).
\eea
Hence the correlation function $C_{12} = \cos \omega t $ (for $t_1 = 0$, $ t_2 = t$), may be read off from the measured probability, assuming that $\lambda$ is known, explicitly
\beq
C_{12} = 1 - \frac { p(1) }{2 \lambda^2}
\eeq
Since $\lambda$ is generally taken to be small, any imprecision in its value could have a large effect on the resulting value of $C_{12}$, hence $\lambda$ will need to be known quite precisely.
(This feature commonly arises in weak measurements of the correlation function. See, for example Ref.\cite{Gog}).
Similarly we find
that the probability of no detection is
\bea
p(0)  &=& \langle \psi | \hat A_0^2 (t) | \psi \rangle,
\nonumber \\
&\approx& 1 - 2 \lambda^2 (1 - \cos \omega t),
\nonumber \\
&=& 1 - 2 \lambda^2 (1 - C_{12}).
\eea

Note that, as anticipated, these probabilities for the final ancilla states do not give direct measurements of $p(S)$ and $p(D)$,
Eqs.(\ref{same}), (\ref{diff}) (unlike the case in Ref.\cite{Halprev}). In fact, the ancilla probabilities may be written
\bea
p(1) &=& 4 \lambda^2 p (D)
\\
p(0) &=& p(S) + ( 1 - 4 \lambda^2) p(D)
\eea
The reason for the difference between $p(0), p(1)$ and $p(S)$, $p(D)$ is that the detector only detects a fraction $4 \lambda^2 $ of histories in which $Q$ changes sign, not all of them. Hence $p(0)$ may be interpreted as the probabilities for those histories in which $Q$ did not change sign, plus those that did change sign but did not trigger the ancilla. Although note that $p(0)$ still arguably corresponds to histories in which there is no interaction (and we check this explicitly below).

Consider now the issue of the invasiveness of the measurement. To what degree is the system state affected by the interaction? Tracing out the ancilla, the system reduced density operator is
\beq
\rho = \hat A_0 (t) | \psi \rangle \langle \psi |\hat A_0(t) +\hat A_1 (t) | \psi \rangle \langle \psi | \hat A_1(t)
\eeq
An interesting choice to make here is a maximally mixed initial state for the system, so $ | \psi \rangle \langle \psi |$ is replaced with $\rho_0 = \half \id $ and we immediately see that $\rho_t = \rho_0$, so the system density operator is undisturbed by the interaction. However, this is potentially misleading -- it simply means that there is no average disturbance (but this still leaves opportunity for hidden variable explanations by classical models with disturbing measurements, as noted in Refs.\cite{SOS,KGBB,HalQ,KoBr}).

The more relevant quantity is the disturbance to the system conditional on a given ancilla state. This conditional disturbance
allows a natural comparison with an ideal negative measurement, since there we expect no disturbance only for certain detector outcomes but not for all outcomes.
Consider therefore the average of any system operator ${\mathcal O}_S$
conditional on the ancilla remaining in its initial state, i.e. conditional on no detection,
\beq
\langle {\mathcal O}_S \rangle_{|0\rangle} = \frac{ \langle \psi | \hat A_0 (t) {\mathcal O}_S \hat A_0(t) | \psi \rangle } { p(0) }.
\eeq
This clearly in general depends on the interaction so there is a disturbance. However, at this point we note that we are free to choose the initial system state, since the correlation function we seek does not depend on it. In particular, we can choose $ |\psi \rangle$ to be an eigenstate of $\hat A_0(t)$, that is, of $\sigma_x$. We thus have
\beq
\hat A_0 (t) | \psi \rangle = \langle \hat A_0 (t) \rangle | \psi \rangle,
\eeq
and also $p(0) = \langle \hat A_0^2 (t) \rangle = \langle \hat A_0 (t) \rangle^2 $ and it immediately follows that
\beq
\langle {\mathcal O}_S \rangle_{|0\rangle}  =\langle \psi | {\mathcal O}_S | \psi \rangle
\label{nodis}
\eeq
Hence there is no disturbance to the system at any time if the ancilla is found in the $|0 \rangle $ state.
Differently put, the part of the system state entangled with $|0 \rangle$ in Eq.(\ref{Psit}), is unchanged by the interaction, except for multiplication by a $c$-number.
This situation is clearly the quantum analogue of the classical case in which
$ Q(t_2) = Q(t_1)$, i.e. the analogue of an ideal negative measurement, since we expect the system to be unchanged in this case. As noted earlier, there are no quantum states in which this holds exactly in the quantum case, but it does hold in the average for some states, for example a $\sigma_x$ eigenstate, the choice made here. In general, classical arguments for non-invasiveness in LG tests do not persist to the quantum level, but here we see that the corresponding quantum situation is in fact non-invasive for a suitable choice of initial state.

Another possibility is to condition instead on the ancilla state $ | 1 \rangle $ and examine the conditional average,
\beq
\langle {\mathcal O}_S \rangle_{|1\rangle} = \frac{ \langle \psi | \hat A_1 (t) {\mathcal O}_S \hat A_1(t) | \psi \rangle } { p(1) }.
\eeq
In this case, it is natural to choose the initial state to be an eigenstate of $\hat A_1 (t)$, that is, of $\sigma_y$, and we again get a result of the form Eq.(\ref{nodis}), so again the system state is undisturbed by the measurement.
This choice actually corresponds to the classical situation with a definite but non-zero value of $ Q(t_2) - Q(t_1)$. To see this, note that in Eq.(\ref{Smat}), which shows that the ancilla couples to $ \hat Q(t_2) - \hat Q(t_1)$, this expression refers to the interaction picture, so when acting on states, we need to consider the expression,
\beq
e^{ - i H_S t} \left( \hat Q(t) - \hat Q(0) \right) = \hat Q e^{- i H_S t} - e^{ - i H_S t} \hat Q
\eeq
where $H_S$ is the system Hamiltonian. With a little algebra it is easily seen that the right-hand side is proportional to $\sigma_y$. 
However, referring back to the classical argument for non-invasiveness in this case, which relies on the fact that the future evolution of the system is irrelevant once $Q$ has changed sign, there is no obvious need in this case to have an initial state which is undisturbed by the interaction.

A perhaps more striking form of non-invasiveness was noted in the similar model in Ref.\cite{Halprev}, in which for certain system initial states there was complete disentanglement of system and ancilla, with no detectable effect on the system state.
This does not seem to be possible in this model.

Finally, we may use this quantum model to assess whether the back reaction of the ancilla on the primary system may cause the final value of the ancilla state to revert to the undetected state $|0 \rangle$, instead of remaining in the detected state $| 1 \rangle $ after a sign change of $Q$.
We have shown (see Eq.(\ref{xi})) that probability for the primary system alone to perform two sign changes is small in general for short timesclales. Now, armed with a particular measurement model, we may ask a similar question about the behaviour of the ancilla states.
We thus consider the total system amplitude for the ancilla to follow the history 
$ |0\rangle \rightarrow |1 \rangle \rightarrow |0\rangle$ at times $0$, $t$, $2t$:
\beq
| \Psi_{010} \rangle = \id \otimes |0 \rangle \langle 0 | \ e^{ - i H t} 
\  \id \otimes |1 \rangle \langle 1 | \ e^{ - i H t} \ |\psi \rangle \otimes |0 \rangle 
\eeq
where $H$ denotes the total system Hamiltonian, Eq.(\ref{ham}). Using Eq.(\ref{Psit}) (and similar relations with initial state $| 1 \rangle $) it is readily shown that
\beq
 | \Psi_{010} \rangle = \hat A_1 (t)^2 |\psi \rangle \otimes | 0 \rangle
\eeq
and hence the probability is
\beq
p_{010} = \frac {16 \lambda^4 \omega^4 } { \Omega^4} \sin^4 \left( \frac { \Omega t} {2} \right)
\eeq
For comparison the amplitude for the history  $ |0\rangle \rightarrow |1 \rangle \rightarrow |1\rangle$ is
\beq
|\Psi_{011} \rangle =  \hat A_0 (t) \hat A_1 (t) | \psi \rangle \otimes | 1 \rangle
\eeq
with probability
\beq
p_{011} = \frac {4 \lambda^2 \omega^2 } { \Omega^2} \sin^2 \left( \frac { \Omega t} {2} \right)
\left[ \cos^2  \left( \frac { \Omega t} {2} \right) + \frac { \omega^2 } { \Omega^2} \sin^2  \left( \frac { \Omega t} {2} \right) \right]
\eeq
Hence $p_{010}$ is easily seen to be smaller than $p_{011}$ by a factor of $\lambda^2$ for small $\lambda$. 
Thus for $\lambda \ll 1$ the probability of an incorrect final ancilla reading due to the back action of the ancilla on the primary system is small.

The correlation function we seek arises in the measured probabilities $p(0)$ and $p(1)$ as a small correction of order $\lambda^2$ about the zero interaction result. However, the measurement disturbance  does not kick in until order $\lambda^4$, so this is the precise sense in which the measurement in the case of no sign changes of $Q$ is approximately non-invasive for small $\lambda$. This therefore represents an improvement on the usual weak measurement approach where the disturbance is very small but the small bias about no disturbance from which the desired physical result is obtained is of the same order.

\section{Summary and Conclusion}

We have presented a protocol for non-invasive measurement of the temporal correlation functions of the type appearing in the LG inequalities. It is similar in spirit to the earlier work Ref.\cite{Halprev}, in that it involves direct measurement of the two types of histories for the system, in which $Q$ has either the same or different signs at the initial and final times. It was shown that the correlation function can be simply expressed in terms of the time average of the velocity $v = \dot Q $ and that this average can be measured using a weak coupling to an ancilla. Under the assumption that $Q$ changes sign at most once for reasonably short times, for which justification was given, the interactions with the ancilla  in each run are then limited to a single interaction at the moment of time when $Q$ changes sign, or no interaction if there is no sign change.
We argued that these interactions have negligible effect on the future behaviour of the ancilla if the back action is small, which is the case for sufficiently small $\lambda$ as was shown explicitly in a quantum model. We also noted that
an irreverisble ancilla would have the same effect but this was not treated in detail. 
Hence, we found that the measurement is non-invasive in the sense that the measurement disturbance is a factor $\lambda^2$  smaller than the (already small) effect being measured, thereby improving on the situation with standard weak measurements. Furthermore, for runs where $Q$ does not change sign, we noted that the non-invasiveness can be checked experimentally by measuring the primary system final state and we showed that this non-invasiveness can be maintained in a quantum model with a suitable choice of initial state.

\section{Acknowledgements}

I am grateful to Clive Emary, George Knee, Owen Maroney and James Yearsley for useful conversations about the Leggett-Garg inequalities. I would also like to thank George Knee for a critical reading of the first version of this manucript.

\bibliography{apssamp}

\end{document}